\documentclass[a4paper]{article}
\usepackage[utf8]{inputenc}
\usepackage{INTERSPEECH2020}
\usepackage{amsmath}
\usepackage{float}
\usepackage{multirow}
\usepackage{multicol}
\usepackage{bigdelim}
\usepackage{hyperref}
\usepackage{stfloats}
\usepackage{adjustbox}
\usepackage{flushend}
\usepackage{blindtext}
\usepackage{cite}

\newlength{\bibitemsep}
\newlength{\bibparskip}
\let\oldthebibliography\thebibliography
\renewcommand\thebibliography[1]{%
  \oldthebibliography{#1}%
  \setlength{\parskip}{\bibitemsep}%
  \setlength{\itemsep}{\bibparskip}%
}

\title{LT-LM: a novel non-autoregressive language model for single-shot lattice rescoring}
\name{Anton Mitrofanov$^{1,2}$, Mariya Korenevskaya$^2$, Ivan Podluzhny$^{1,2}$, Yuri Khokhlov$^2$, Aleksandr Laptev$^1$, Andrei Andrusenko$^1$, Aleksei Ilin$^2$, Maxim Korenevsky$^2$, Ivan Medennikov$^{1,2}$, Aleksei Romanenko $^{1,2}$}
\address{
  $^1$ITMO University, St. Petersburg, Russia\\
  $^2$STC-innovations Ltd, St. Petersburg, Russia}
\email{\{mitrofanov-aa, korenevskaya, podluzhnyi, khokhlov\}@speechpro.com, aalaptev@itmo.ru, \{andrusenko, ilin-a, korenevsky, medennikov, romanenko\}@speechpro.com}

\begin{document}

\maketitle

\begin{abstract}
Neural network-based language models are commonly used in rescoring approaches to improve the quality of modern automatic speech recognition (ASR) systems. Most of the existing methods are computationally expensive since they use autoregressive language models. We propose a novel rescoring approach, which processes the entire lattice in a single call to the model. The key feature of our rescoring policy is a novel non-autoregressive Lattice Transformer Language Model (LT-LM). This model takes the whole lattice as an input and predicts a new language score for each arc.
Additionally, we propose the artificial lattices generation approach to incorporate a large amount of text data in the LT-LM training process. 
Our single-shot rescoring performs orders of magnitude faster than other rescoring methods in our experiments. It is more than 300 times faster than pruned RNNLM lattice rescoring and N-best rescoring while slightly inferior in terms of WER.

\end{abstract}

\section{Introduction}
Language models (LMs) are a significant part of automatic speech recognition (ASR) systems. They can reduce the overall system error rate \cite{Hori2007Efficient, Han-2018-TheC2, Luscher-2019-RWTH, Toshniwal-2018-ACO} and make the recognition results more human-readable~\cite{liao2021generating}. 
Back-off n-gram LMs \cite{Goodman_2001} are traditionally used in hybrid ASR systems \cite{Mohri2002Weighted}, but the Neural Network Language Models (NNLMs) outperform n-gram ones \cite{Mikolov2010RecurrentNN, Medennikov-2016, Liu_2016, stc_chime6, Medennikov2019TheSA, Saon2017_eng} due to their ability to capture longer context and extract more meaningful information from text.
However, NNLMs are computationally expensive, and their usage in ASR systems is challenging. The most popular way to apply them is the two-pass method\cite{Sundermeyer-2014-LatticeDA}. First, decoding pass is performed with an n-gram LM. Then, a NNLM can be additionally used to rescore\cite{Liu_2016} the results of the first pass.

There are several widely used rescoring approaches. The simplest one is the N-best list rescoring \cite{deoras-etal-2011-fast}. It is applied separately to each hypothesis of the first pass, ranking them with respect to updated scores. N-best rescoring provides better Word Error Rate (WER) reduction when it operates with a larger number of hypotheses $N$ to re-rank. Usually, hypotheses differ from each other only in a small number of words. Since N-best rescoring does not take this fact into account, it requires a lot of unnecessary computations, which makes this method very computationally expensive. This flaw of N-best rescoring is mitigated when using rescoring approaches dealing with word lattices. There are approaches such as pruned RNNLM lattice rescoring \cite{Xu_2018}, fast N-best rescoring \cite{Beck_2020} and parallelizable lattice rescoring\cite{li2021parallelizable} which exploit the similarity of hypotheses to optimize computations.
Nevertheless, these rescoring approaches are rather slow and computationally expensive as they require multiple LM calls.

In this article, we introduce a new non-autoregressive language model that processes the whole lattice in a single shot during the rescoring process. It is trained directly on ASR lattices, unlike conventional LMs trained on text sequences. We used the Transformer architecture~\cite{Vaswani_2017} for this purpose because it can be easily adapted for training on lattices~\cite{sperber-etal-2019-self, shiv-2019-novel, li-etal-2020-flat} and named this special model Lattice Transformer Language Model (LT-LM). This model takes a whole lattice as input and outputs a language weight for each arc. Using such a model in a rescoring process makes it possible to re-rank the lattice hypotheses in a single call to the model. This makes the rescoring process significantly faster and reduces the number of computational resources compared to rescoring with autoregressive models.

Autoregressive LMs are usually trained on large amounts of text data, which helps them to provide significant WER reduction in the rescoring task. Since LT-LM is supposed to be trained on ASR results and corresponding speech transcripts, it is challenging to obtain enough training data. To cope with the lack of training data, we developed a method of artificial lattices generation from large amounts of texts, using just a small subset of labeled audio data (about 1-2 hours). This made it possible to significantly increase the amount of training data for LT-LM and obtain rescoring results close to ones corresponding to modern autoregressive LMs.


\section{Rescoring Task}
\label{sec:rescoring}

Normally, the decoding process searches for the best ASR hypothesis according to the Maximum A Posteriori (MAP) rule:
\begin{equation} \label{eq:MAPrule}
\begin{split}
W^* &= \underset{W}{\mathrm{argmax}} \, p(X|W)^a P(W)^l,        
\end{split}
\end{equation}
where $W$ stands for a recognition hypothesis, $W^*$ is the best hypothesis, $a$ and $l$ are adjustable acoustic and language weights, respectively, while $p(X|W)$ is an acoustic score, and $P(W)$ is a language score.

The process of searching for the best hypothesis requires applying the language model to all possible word sequences to estimate their language probabilities. This process is fast when n-gram LMs are used. Typically, NNLMs work much slower but significantly outperform n-gram ones in terms of WER. Thus, it is necessary to reduce the number of recognition hypotheses to apply neural network-based LMs. It can be done with a two-pass approach. First, decoding with an n-gram LM is performed, and a list of hypotheses with the best scores is saved. The result of the first pass can be compactly represented as a weighted directed acyclic graph called a lattice. Lattice arcs are usually supplied with word labels, as well as language and acoustic scores. Lattice always has a single initial state and at least one final state.

The second part of the two-pass method is the rescoring process. It updates language scores using neural network-based LM and reorders hypotheses. Generally, after such a procedure, the best hypothesis changes to the one containing fewer errors. The score of the best hypothesis after rescoring can be evaluated as follows:
\begin{equation}
\label{eq:MAProolRescoring}
W^* = \underset{W \in \mathrm{lattice}}{\mathrm{argmax}} P(X|W)^a P_1(W)^{l_1} P_2(W)^{l_2},
\end{equation}
where $P_1(W)$ and $P_2(W)$ are the probabilities of n-gram LM and rescoring LM, respectively, and $l_1,l_2$ are corresponding LM weights.

The language model estimates a probability of a given sequence of words. However, the rescoring process operates only with the most probable hypotheses of the first pass. In order to take this fact into account, we propose a special LM, which estimates the language scores directly for a set of lattice hypotheses. We named it Lattice Transformer Language Model and trained it on lattices instead of text data.

\section{Lattice Transformer Language Model}
\label{sec:lt}

While the text is a sequence of words, a lattice is a non-linear structure. This means that the order of words in a lattice is determined by lattice topology. The Transformer architecture can be naturally adapted to training on lattices\cite{sperber-etal-2019-self}, in contrast to, for example, recurrent neural networks, for which the order of the input word sequence is essential. Such adaptation is possible due to the self-attention component of the Transformer architecture. This module does not depend on the order of its elements because it computes a weighted sum of the input sequence. Transformer uses positional encoding to introduce the information about the sequence order into its latent representations. The following subsection describes the proposed approach of lattice topology encoding for use in Transformer.

\subsection{Positional Encoding}

One of the most common ways to encode positional information is combining embeddings of the input elements with the so-called positional embeddings \cite{Gehring_2017}. Positional embeddings are usually obtained from the embedding matrix trained together with the model. Each embedding corresponds to a specific position in the input sequence. This encoding method can be easily adapted for training on data structures different from sequences (e.g., lattices).

We propose the following pipeline for positional encoding of lattices. The first step is representing the lattice as a set of its arcs. Then, we propose adding several auxiliary arcs to keep information about the initial and final states of a lattice. They serve the same purpose as the SOS and EOS tokens in the case of text-based learning. An auxiliary arc with the symbol $<$s$>$ comes to the initial state, and auxiliary arcs with the symbol $<$/s$>$ come from each final state. Here symbols $<$s$>$ and $<$/s$>$ correspond to the beginning and the end of the sentence, respectively. The next step is the topological sort of lattice to ensure the proper order of its states.

The position of the arc $(x,y)$ in the lattice is determined by the indices of its source and destination states $x$ and $y$. Thus, it is possible to encode the topological information of each arc in the lattice using the sum of positional embeddings of its source and destination states, respectively. Two embedding matrices for source and destination states are trained separately to account for arcs direction. Finally, the total arc embedding is obtained by adding its positional embedding to a corresponding word embedding.

\subsection{Targets and loss}

The hypothesis with the least WER in the lattice is called the oracle hypothesis. The ideal rescoring would provide this hypothesis as a result. For this reason, LT-LM is trained to find the oracle hypothesis in the lattice. This is done in the following way.
First, the oracle path is obtained from each lattice prepared as described above. If there are several oracle paths in lattice, the target path is chosen randomly. Then lattice arcs are split into two classes. All arcs comprising the oracle path are assigned to class 1, and the others --- to class 0.  LT-LM is trained to minimize the binary cross-entropy for each arc.

\subsection{Single-shot Rescoring with LT-LM}

The LT-LM key feature is the ability to generate probabilities for all arcs in the lattice in a single call to the model. LT-LM predicts the probability for each arc of the lattice to belong to the oracle path. These probabilities can be used in combination with original LM and AM scores to generate the final rescoring result. Then, the best hypothesis is chosen using the standard forward algorithm accordingly to (\ref{eq:MAProolRescoring}). The described method is significantly faster than other rescoring methods requiring several calls to the model. This fact is discussed in detail in subsection \ref{subsec:speedup}.

\begin{figure*}[ht]
    \vspace{-20pt} 
    \centering
    \includegraphics[width=0.9\textwidth]{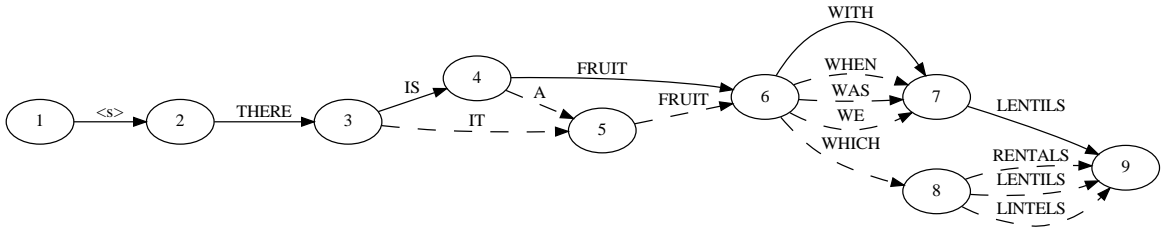}
    \caption{A beginning part of an artificial lattice. The whole lattice contains 53 arcs, while the reference text contains 22 words.}
    \label{fig:genlat}
    \vspace{-10pt} 
\end{figure*}

\section{Lattices generation from text}
\label{sec:latgen}

As mentioned above, recognition lattices and ground-truth word-level transcriptions are needed for LT-LM training. At the same time, classic LMs are trained in an unsupervised manner using large amounts of text data, which are much easier to collect. In order to utilize large amounts of available text data for LT-LM training, we propose an algorithm that generates realistic artificial lattices directly from the text.

It is necessary to imitate the acoustic similarity of words to generate realistic lattices from the text. We propose the following approach for generating lattices from text data as close as possible to real ones:
\begin{itemize}
    \item simulate a per-frame alignment from text without using audio data;
    \item simulate a probability distribution over acoustic classes for each frame of the alignment;
    \item decode the resulting sequence with a beam search algorithm to generate lattices.
\end{itemize}

We used an acoustic model and a small labeled audio dataset to generate lattices. We estimated the per-frame distribution of acoustic classes of this dataset and built a per-frame alignment. For each acoustic class, we estimated the distribution of its duration in the alignment. We evaluated this distribution from the alignment and calculated the average acoustic model output for each acoustic class. 
Finally, we obtained a matrix of the shape $(A, A)$, where $A$ stands for the number of acoustic classes, and referred it to as the Fake Acoustic Model (FAM).
The $i$-th row in this matrix contains the average distribution of acoustic classes over all frames where the alignment value is equal to $i$.

Alignment can be obtained from the text as follows. Similar to the forced alignment pipeline, each sentence is converted to a graph containing all possible representations of the corresponding word sequence in the form of acoustic classes. If there is an audio corresponding to the text, the most probable alignment can be obtained using the Viterbi algorithm. In contrast, we propose to sample a random path from the alignment graph. This path should not contain the same acoustic classes on adjacent arcs. Unlike the real alignment, the obtained paths are time-agnostic. Thus, it is necessary to introduce the time component and ``stretch'' this path to simulate realistic alignment. We sampled a random number $N$ from the estimated distribution for each acoustic class and stretched this class by $N$ frames. We referred to the resulting sequence of acoustic classes as the Fake Alignment (FAli).

Then, artificial lattices can be generated by performing the conventional decoding procedure on FAli. The probability distribution of acoustic classes for each frame is sampled from the FAM model.
Finally, the standard decoding procedure can be applied to generate artificial lattices.

Figure \ref{fig:genlat} represents a part of generated lattice for an utterance from \textit{dev\_clean} subset of LibriSpeech corpus. Solid lines mark the path corresponding to the oracle hypothesis, while dashed ones indicate arcs of alternative hypotheses added during the lattice generation. 
The structure of a simulated lattice is similar to the structure of a real one. The words from alternative hypotheses of generated lattice sound similarly to corresponding ones from the oracle path.

\section{Experiments}
\label{sec:experiments}

\subsection{Baselines}
Our experiments with LT-LM were performed using the LibriSpeech corpus \cite{librispeech}, which contains 960 hours of English audiobooks for AM training and 5Gb of extra text data for LM training. The Kaldi \cite{Povey_ASRU2011} recipe for LibriSpeech\footnote{\href{https://github.com/kaldi-asr/kaldi/blob/master/egs/librispeech/s5/local/chain/tuning/run_tdnn_1d.sh}{kaldi/egs/librispeech/s5/local/chain/tuning/run\_tdnn\_1d.sh}} was used for the baseline AM training. The baseline WER results were obtained using the large 3-gram LM from this recipe. As rescoring baselines, we used a Transformer-LM\footnote{\href{https://github.com/espnet/espnet/blob/master/egs2/librispeech/asr1/conf/tuning/train_lm_transformer2.yaml}{espnet/egs2/librispeech/asr1/conf/tuning/train\_lm\_transformer2.yaml}} as well as LSTM-LM\footnote{\href{https://github.com/espnet/espnet/blob/master/egs2/librispeech/asr1/conf/tuning/train_lm_adam.yaml}{espnet/egs2/librispeech/asr1/conf/tuning/train\_lm\_adam.yaml}} from ESPNet framework \cite{watanabe2018espnet}. Transformer-LM consists of 16 transformer blocks. Each block has 8 attention heads of dimensions 512 and 2048 units in feed-forward layers. LSTM-LM has 4 layers of 2048 neurons each. The target texts were tokenized into 5000 tokens by Byte Pair Encoding (BPE) algorithm~\cite{sennrich-2016-neural} from the SentencePiece tokenization toolkit~\cite{kudo2018}. The total number of parameters of LSTM and Transformer LMs are 155M and 54M, respectively. We considered two rescoring approaches as baselines: N-best rescoring and pruned RNNLM lattice rescoring~\cite{Xu_2018}. The last one was performed with 4-gram approximation and the beam of 6.

\begin{table}[!b]
    \vspace{-10pt} 
    \begin{adjustbox}{width=1\linewidth}
    \begin{tabular}{c|c|c|c|c|c}
    \toprule
         \multirow[b]{2}{*}{LM} & \multirow[b]{2}{*}{method} & \multicolumn{2}{c|}{dev} & \multicolumn{2}{c}{test} \\ [\smallskipamount]
         \cline{3-6} \rule{0pt}{10pt}
          & & clean & other & clean & other \\
        \midrule
        \multicolumn{6}{c}{lattices pruned with beam 8} \\
        \midrule
         3-gram & - & 3.41 & 9.26 & 3.90 & 9.25 \\
         LSTM & 50-best &  2.78 & 7.51 & 2.84 & 7.67 \\
         LSTM & 500-best & 2.56 & 7.55 & 2.86 & 7.67 \\
         LSTM & Pruned & 2.63 & 7.10 & 2.95 & 7.38 \\
         Transformer & 50-best & 2.47 & 8.10 & 2.82 & 8.18 \\
         Transformer & 500-best & \textbf{2.26} & \textbf{7.06} & \textbf{2.63} & \textbf{7.23} \\
         \midrule
         \multicolumn{6}{c}{lattices pruned with beam 4} \\
         \midrule
         LSTM & 50-best & 2.88 & 8.61 & 3.14 & 8.69 \\
         LSTM & 500-best & 2.74 & 7.89 & 3.05 & 8.05	 \\
         Transformer & 50-best & 2.61 & 8.28 & 2.99 & 8.37 \\
         Transformer & 500-best & \textbf{2.50} & 7.51 & \textbf{2.84} & \textbf{7.67} \\
         LT-LM & Single-Shot & 2.51 & \textbf{7.47} & 3.01 & 7.74 \\
         
    \bottomrule
    \end{tabular}
    \end{adjustbox}
    \caption{Rescoring results in WER on LibriSpeech development and test sets for different NNLMs and rescoring methods. ``Pruned'' stands for pruned RNNLM lattice rescoring.}
    \vspace{-10pt} 
    \label{tab:rescoring}
\end{table}

\subsection{LT-LM training}

Initially, we decoded the entire LibriSpeech training set to obtain lattices for the LT-LM training. The model trained using only these lattices did not show the competitive rescoring results due to the following reasons. First, the obtained amount of data was insufficient for the LT-LM training and led to model overfitting. Second, the word diversity of the training utterances texts did not cover the vocabulary of the baseline n-gram LM trained with additional text data.
We used a small subset of training data of 2 hours and additional LibriSpeech texts for the artificial lattices generation to overcome these problems. 
Decoding is the most computationally intensive part of the generation process. Due to the resource limitation, we accelerated the decoding by reducing the beam from 16 to 13 and the number of active hypotheses from 5000 to 3000. Finally, we pruned lattices with beam 4 to speed up the training process. 
LT-LM was trained on word lattices, and the overall vocabulary consisted of 200K words. 

We used the Fairseq framework~\cite{ott2019fairseq} for LT-LM training. The model had 8 transformer encoder layers, each of 8 attention heads with 816 embedded dimensions and 2048 feed-forward units. The total number of the model parameters was 211M. LT-LM was trained using Adam optimizer with a linear learning rate warmup for the first 4000 iterations and with a constant learning rate of 3e-5 after warmup. The use of gradient checkpointing~\cite{Chen2016TrainingDN} significantly reduced the memory usage and allowed us to increase the batch size from 4 to 64.
Our best LT-LM was trained for six epochs on a combination of real and artificial lattices.

\subsection{Rescoring WER performance}

The rescoring results for all models mentioned above are presented in Table~\ref{tab:rescoring}. The first row of the table demonstrates the rescoring results of the best 3-gram LM from the LibriSpeech Kaldi recipe. The next three rows correspond to the LSTM-LM results: 50-best and 500-best rescoring and pruned RNNLM lattice rescoring~\cite{Xu_2018}, respectively. The next two rows correspond to 50-best and 500-best rescoring with Transformer-LM and the latest one shows the best WER results on LibriSpeech development and test sets. 
The NNLM weight of 0.8 was used for all the described approaches.
Since LT-LM was trained on lattices pruned with a beam of 4, we used the same beam value in test time to avoid the mismatch between training and evaluation. Also, we provide the results of N-best rescoring with a beam equal to 4. These results are demonstrated in the second part of Table~\ref{tab:rescoring} and reflect the influence of beam value on the rescoring quality.
As the final result, the single-shot approach with LT-LM shows the quality comparable to N-best rescoring with LSTM-LM.

\begin{table}[]
    \vspace{0pt} 
    \begin{adjustbox}{width=1\linewidth}
    \begin{tabular}{c|c|c|c|c|c}
    \toprule
         \multirow[b]{2}{*}{LM} & \multirow[b]{2}{*}{method} & \multicolumn{2}{c|}{dev} & \multicolumn{2}{c}{test}  \\
         [\smallskipamount]
         \cline{3-6} \rule{0pt}{10pt}
          & & clean & other & clean & other \\ 
    \midrule
        \multicolumn{6}{c}{lattices pruned with beam 8} \\
    \midrule
         LSTM & 50-best & 12m25s & 13m8s & 12m32s & 13m21s \\
         LSTM & 500-best & 48m22s & 59m26s & 49m5s & 60m39s \\
         LSTM & Pruned & 13m47s & 34m & 14m23s & 35m38s \\
         Transformer & 50-best & 10m25s & 11m48s & 10m25s & 11m59s \\
         Transformer & 500-best & 41m27s & 55m42s & 39m8s & 54m47s \\
     \midrule
         \multicolumn{6}{c}{lattices pruned with beam 4} \\
     \midrule
         LSTM & 50-best & 9m15s & 10m27s & 9m6s & 11m6s \\
         LSTM & 500-best & 19m13s & 32m & 17m58s & 32m34s \\
         Transformer & 50-best & 8m11s & 9m43s & 7m59s & 9m43s \\
         Transformer & 500-best & 15m41s & 27m55s & 15m33s & 28m44s \\
         LT-LM & Single-Shot & \textbf{3.8s} & \textbf{4.5s} & \textbf{3.7s} & \textbf{4.6s} \\
    \bottomrule
    \end{tabular}
    \end{adjustbox}
    \caption{Rescoring GPU running time in minutes and seconds}
    \vspace{-20pt} 
    \label{tab:rescoring_time}
\end{table}

\subsection{Rescoring speedup}
\label{subsec:speedup}

Rescoring was performed on a single Nvidia GeForce RTX 2080 ti GPU. We measured the GPU running time of different rescoring approaches on the LibriSpeech development and test sets to compare their speed performances. 
The GPU running time results for all rescoring approaches except 500-best were averaged over ten launches. The results for 500-best rescoring were averaged over only two launches due to high computational cost per run.
The rescoring for both LT-LM and N-best lists was performed with a batch size equal to 64. The results are presented in Table~\ref{tab:rescoring_time}.  It can be seen that the single-shot rescoring with LT-LM was orders of magnitude faster than the other methods. For example, LT-LM rescoring is about 350 times faster than pruned RNNLM lattice rescoring, while having comparable quality. At the same time, Transformer-LM 500-best rescoring outperforms LT-LM rescoring for a maximum of 0.5\% WER on test\_other, but LT-LM performs more than 630 times faster!
The reasons for such acceleration can be explained as follows.

The main advantage of LT-LM is that the rescoring of the entire lattice is performed in a single-shot manner. 
This leads to the significant speedup of the rescoring process. Table~\ref{tab:avg_lens} demonstrates the statistics obtained for the LibriSpeech \textit{dev\_clean} and \textit{dev\_other} sets. The left half of the table shows the average length of sequences processed for a single LM call, while the right half shows the number of such calls.

In the case of N-best rescoring, the length of the processed sequence equals the number of words in the recognition hypothesis. For the pruned RNNLM lattice rescoring, the length of the processed word sequence always equals one, because the algorithm processes arcs in the lattice one by one. Finally, since LT-LM processes the whole lattice, the corresponding word sequence length during the rescoring process is equal to an overall number of arcs in the lattice. 
Respectively, the number of LM calls is equal to the total amount of processed hypotheses for the N-best rescoring, to the total amount of arcs involved in rescoring in the composed lattice for the pruned RNLLM rescoring, and to the number of lattices for the LT-LM one.

At first glance, it looks surprising that the average amount of arcs in lattices is close to the average N-best hypothesis length (or even less for the case of 500-best).
The matter of things is as follows.
The N-best list for the short utterances usually contains only a few hypotheses (much less than N). These hypotheses are mostly rather short, and the number of arcs in such lattices is only slightly larger than the hypotheses lengths.
The number of arcs in lattices corresponding to long utterances is usually larger than the hypotheses length by a moderate factor.
But the N-best lists corresponding to such lattices contain a lot of long hypotheses to be rescored.
Thus, the N-best rescoring needs to process significantly fewer short hypotheses than long ones.
As a result, the total average length of processed hypotheses is biased towards the lengths of long hypotheses from large lattices. However, an average number of arcs in lattices is not subject to this bias.

According to Table~\ref{tab:avg_lens}, there is no significant difference in the average sequence lengths for LT-LM and N-best rescoring methods.
However, the number of model calls to process for LT-LM is several orders of magnitude less than for all other rescoring methods presented in the table.

\begin{table}[]
    \vspace{0pt} 
    \centering
    \begin{tabular}{c|c|c|c|c}
    \toprule
         \multirow[b]{2}{*} {method} & \multicolumn{2}{c|}{av. seq. len.} & \multicolumn{2}{c}{num. of model calls} \\ [\smallskipamount] \cline{2-5} \rule{0pt}{10pt}
          & clean & other & clean & other \\
    \midrule
         50-best & 23 & 19 & 105K & 127K \\
         500-best & 27 & 21 & 795K & 1.1M \\
         Pruned & 1 & 1 & 196K & 456K \\
         Single-Shot & 26 & 33 & 2703 & 2864 \\
    \bottomrule
    \end{tabular}
    \caption{Average word sequence lengths and numbers of model calls for different rescoring methods on LibriSpeech dev\_clean and dev\_other sets.}
    \vspace{-25pt} 
    \label{tab:avg_lens}
\end{table}

\section{Conclusion and future work}
\label{sec:conclusion}
    
We presented a novel single-shot approach to rescoring of ASR results. This approach allows to rescore the entire lattice in a single LM call.
While comparable in quality to other popular rescoring approaches, the proposed one is faster by orders of magnitude. The proposed approach utilizes a special Lattice Transformer Language Model, which deals with ASR results presented as lattices.

In the future, we plan to adapt sequence-discriminative losses for LT-LM training. We believe that they can improve the LT-LM quality and help it to outperform autoregressive LMs in terms of WER. It is also interesting to apply LT-LM on far-field speech datasets, which are more challenging for ASR.

\section{Acknowledgements}

This work was partially financially supported by ITMO University.


\nocite{*}

\end{document}